\documentclass[copyright]{eptcs}
\providecommand{\event}{TFPIE} 
\usepackage{breakurl}             
\usepackage{graphicx}

\title{Simple Balanced Binary Search Trees}
\author{Prabhakar Ragde
\institute{Cheriton School of Computer Science\\
University of Waterloo\\
Waterloo, Ontario, Canada}
\email{plragde@uwaterloo.ca}
}
\def\titlerunning{Simple Balanced Binary Search Trees}
\def\authorrunning{P.~Ragde}

\DeclareGraphicsExtensions{.png,.pdf}
\begin{document}
\maketitle

\begin{abstract}
Efficient implementations of sets and maps (dictionaries) are
important in computer science, and balanced binary search trees are
the basis of the best practical implementations. Pedagogically,
however, they are often quite complicated, especially with respect to
deletion. I present complete code (with justification and analysis not
previously available in the literature)
for a purely-functional implementation based on AA trees, which is the
simplest treatment of the subject of which I am aware.
\end{abstract}

\section{Introduction}

Trees are a fundamental data structure, introduced early in most
computer science curricula. They are easily motivated by the need to
store data that is naturally tree-structured (family trees, structured
documents, arithmetic expressions, and programs). We also expose
students to the idea that we can impose tree structure on data that is
not naturally so, in order to implement efficient manipulation
algorithms. The typical first example is the binary search tree.
However, they are problematic.

Naive insertion and deletion are easy to present in a first course
using a functional language (usually the topic is delayed to a second
course if an imperative language is used), but in the worst case, this
implementation degenerates to a list, with linear running time for all
operations. The solution is to balance the tree during operations, so
that a tree with $n$ nodes has height $O(\log n)$. There are many
different ways of doing this, but most are too complicated to present
this early in the curriculum, so they are usually deferred to a later
course on algorithms and data structures, leaving a frustrating gap.
Furthermore, most presentations in conventional textbooks avoid
recursion in favour of loops, use mutation to implement local
rotations, and often, due to space constraints, present final
optimized code rather than intermediate versions that might aid in
understanding. 

A purely functional approach to data structures usually facilitates
educational discussion, and it comes close in this case, but does not
close the gap. The invariants for red-black trees \cite{GS} (an encoding of 2-3-4
trees, which can be viewed as a special case of B-trees) are difficult
to motivate, but the code for maintaining it in the case of insertion
is short and easy to justify, as Okasaki showed \cite{Ok}.

Okasaki's code for insertion into a red-black tree applies a
\texttt{balance} function to each tree constructed with
the result of a recursive call. His compelling contribution was
to notice that the heart of the rebalancing function in the case of
insertion consists of five cases, each one line long and exhibiting a
pleasing symmetry (the right-hand sides of each case can be made
identical). However, he does not handle deletion, with good reason; it
is much more complicated.

Stefan Kahrs added code for deletion from red-black trees \cite{Ka1} that is the basis for
a Haskell \cite{H} library available through Hackage \cite{DSRBT}, but his code
is complex, and comes with no explanation. An explanation of related
code is available in a journal paper \cite{Ka3}, but his goal in this
paper was to use advanced features of the Haskell type system to
enforce invariants, rather than to ease matters for undergraduates.
Matthew Might, in a lengthy blog post, calls code based on Kahrs's
work ported to other languages such as Scala ``Byzantine''\cite{Mi}. 

Might tackles
the topic from a pedagogical point of view, introducing two new
colours (double-black and negative black) during the deletion phase, a
notion of colour arithmetic, customized match expanders for his Racket \cite{R}
implementation, and three code phases (removing, bubbling,
rebalancing). This is probably the best presentation of deletion from
red-black trees, but it is still not good enough to close the gap; the
code is still too long, and the justification too complicated.

The simplest deletion code known (addition is also simple) is for
AA trees, an encoding of 2-3 trees, named after their creator
Arne Andersson. Andersson's work \cite{AA} was published in a
conference with a restriction on page length, and while it is
available on his Web site, there is no journal version or longer exposition.
The paper gives the invariants but does not explain how the code
maintains them. The code is written in Pascal; it does use
recursion (he apologizes for this to his audience)
but it also makes heavy use of mutation.

Here we will work out a purely functional implementation. To the best
of my knowledge, this is the first publication of a purely-functional
treatment of AA trees with complete justification and analysis, and it
is simpler than any of the work cited above.

\section{Invariants}

As mentioned in the previous section, AA trees are an encoding of 2-3
trees \cite{AHU}. A 2-3 tree generalizes a binary tree, a node of which contains
one key $k$, a left subtree with keys less than $k$, and a right
subtree with keys greater than $k$. (We assume unique keys, but it is
easy to adapt the code to handle duplicates.) 2-3 trees allow ternary nodes
containing two keys $k_1, k_2$ and three subtrees: a left subtree with
keys less than $k_1$, a middle subtree with keys between $k_1$ and
$k_2$, and a right subtree with keys greater than $k_2$. In contrast
to the naive elementary implementation of binary search trees, 2-3
trees do not allow nodes with a single child, and all leaves are at
the same depth.

2-3 trees are a good approach to understanding the concepts behind
balanced binary search trees and help to explain the invariants of
both red-black trees and AA trees. However, code implementing them
directly gets complex due to the number of special
cases. Consequently, it makes sense to simulate them by binary search
trees. A node in the 2-3 tree is simulated by one or two binary nodes
in the AA tree.  Instead of a colour at each node (as is the case with
the red-black simulation of 2-3-4 trees), we maintain the level that
the key would be at in the 2-3 tree.  An empty tree has level 0,
leaves have level 1, their parents have level 2, and so on. Besides
simplifying the code, this also aids students in understanding how and
why invariants need to be maintained.

An AA tree is a binary search tree, and so the code for searching is
unchanged from the naive implementation (as is the case for all
balanced binary search tree schemes).
To ensure that an AA tree actually does encode a 2-3 tree, it is
necessary to maintain some other invariants as well.

\clearpage 
In the following diagrams illustrating the invariants, 
a node is labelled with its key, and levels
decrease from top to bottom.

\noindent \textbf{AA1}: The left child of a node $x$
has level one less than $x$.

\includegraphics[scale=0.4]{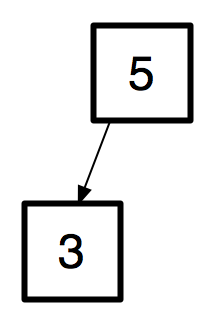}

\noindent \textbf{AA2}: The right child of $x$
has the same level ($x$ is a ``double'' node) or one less ($x$ is a
``single'' node).

\includegraphics[scale=0.4]{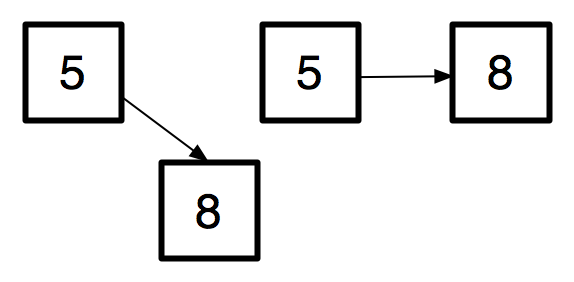}

\noindent \textbf{AA3}: The right child of the right child
of $x$ has level less than $x$.

\includegraphics[scale=0.4]{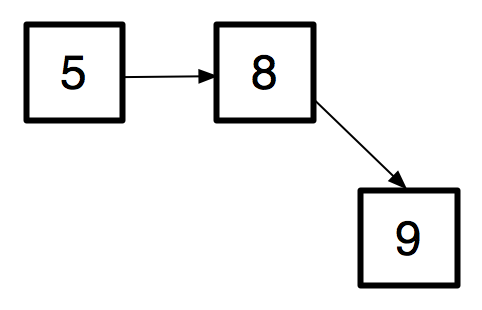}

\noindent \textbf{AA4}: All internal nodes have two children.

The invariants ensure that the height of the AA tree is logarithmic in
the number of nodes. In the 2-3 tree being simulated, all internal
nodes have two or three children.  Internal 2-3 nodes with three
children are simulated by an double AA tree node and its
right child.

\includegraphics[scale=0.4]{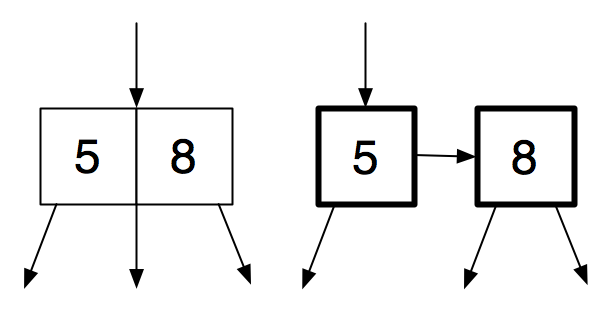}

Leaves of the 2-3 tree may contain one or two values, and all leaves are at
the same depth from the root. This means that the height of
the 2-3 tree is logarithmic in the number of nodes. 
As with red-black trees, a root-node path in the AA tree is at most twice as long
as the corresponding path in the 2-3 tree, thanks to invariant AA3.

\section{Insertion}

For brevity, I will use Haskell in this exposition, though nothing in
the code requires laziness or static typing. A Racket implementation
(Racket is strict and dynamically typed) is also brief compared to the
other options; it is slightly more verbose than the code presented
here due to differences in syntax, not semantics.  We start developing
the code using an elementary implementation of binary search trees
without balancing (and thus no levels to maintain), as would be seen
in a typical introductory course.

\clearpage

In the code and in the diagrams, we will follow the convention that
we will call a tree argument \texttt{t} where possible, and
subtrees will be referred to by letters from the beginning of the
alphabet (\texttt{a}, \texttt{b}, etc.), in lower case as is required
in Haskell. Individual nodes will be referred to by letters from the
end of the alphabet (\texttt{y}, \texttt{z}, etc.). If we need to
pattern-match to gain access to individual fields of the node, we
will use the convention that node \texttt{t} will have level
\texttt{lvt}, left subtree \texttt{lt}, key \texttt{kt}, and right
subtree \texttt{rt}.

\begin{verbatim}
data Bst a = E | T (Bst a) a (Bst a)

-- insertion without balancing
insert :: Ord a => a -> Bst a -> Bst a
insert k E = T E k E
insert k t@(T lt kt rt)
 | k<kt = T (insert k lt) kt rt))
 | k>kt = T lt kt (insert k rt)))
 | otherwise = t
\end{verbatim}

As with Okasaki's code for insertion, we will rebalance each tree
constructed with the result of a recursive application.  We will split
the rebalancing function for AA trees into two two-line helper
functions, \texttt{skew} and \texttt{split}, which will also be useful
for deletion.  (These functions appear in Andersson's paper with the
same names, though in imperative versions that use mutation.)  Note
the introduction of the integer level argument for the \texttt{T} data
constructor.

\begin{verbatim}
data AAt a = E | T Integer (AAt a) a (AAt a)

-- insertion with balancing
insert :: Ord a => a -> AAt a -> AAt a
insert k E = T 1 E k E
insert k t@(T lt a kt b)
 | k<kt = split $ skew $ T lt (insert k a) kt b))
 | k>kt = split $ skew $ T lt a kt (insert k b)))
 | otherwise = t
\end{verbatim}

In general, a subtree produced as the result of
a recursive application of \texttt{insert} may have had the level
of its root increased by one, and means that the tree
constructed from it by adding in the unchanged left or right subtree
may not satisfy the invariants.

If the left child of a node is increased by one,
it is single (as we will see), and
we need the \texttt{skew} transformation
to fix the violation of invariant AA1. (A diagram showing
a pair of trees, such as shown below,
represents ``before'' and ``after'' situations.)

\includegraphics[scale=0.7]{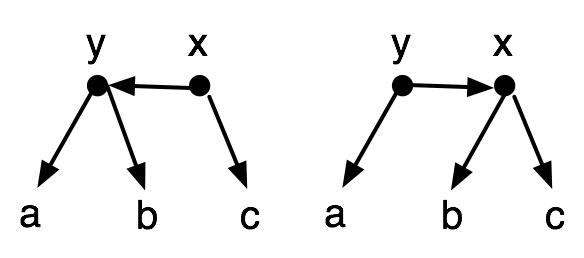}

\begin{verbatim}
skew (T lvx (T lvy a ky b) kx c) | (lvx == lvy) = T lvx a ky (T lvx b kx c)
skew t = t
\end{verbatim}

\clearpage

But the result of skewing could still be a problem if the former root
had a right child at the same level, because invariant AA3 is violated.

\includegraphics[scale=0.7]{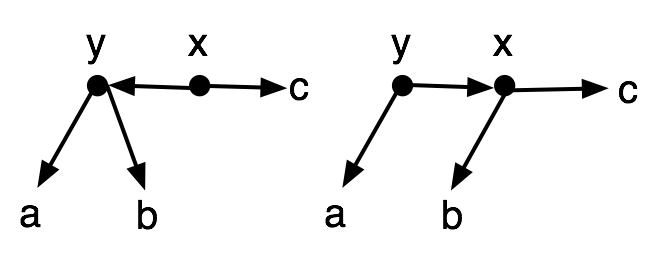}

This is also a problem if the right child of a right child
at the same level has its level raised.
Once again, invariant AA3 is violated after the skew.

\includegraphics[scale=0.7]{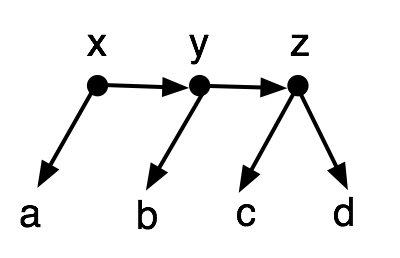}

These situations are handled by the \texttt{split} transformation,
which solves the problem by raising the level of the middle node by one,
ensuring that it is single and the parent of the other two nodes.
All invariants are restored.

\includegraphics[scale=0.6]{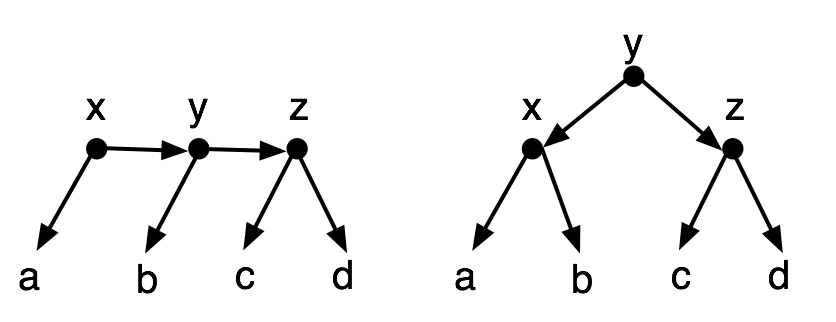}

\begin{verbatim}
split (T lvx a kx (T lvy b ky (T lvz c kz d))) | (lvx == lvy) && (lvy == lvz)
 = T (lvx+1) (T lvx a kx b) ky (T lvx c kz d)
split t = t
\end{verbatim}

Note that it does not hurt to apply \texttt{skew} and \texttt{split}
to trees that do not require those transformations.
This completes the presentation of insertion,
which so far is at least as short and
understandable as Okasaki's, if not more so.
The instructor may choose to stop at this point, 
leaving deletion as a rather challenging exercise.

\section{Deletion}

As with insertion, we start with the code for elementary deletion from
an unbalanced binary search tree. The case that is not straightforward
occurs if the key to be deleted is in an internal node. In this case,
we can, in effect, move the largest key in the left subtree up to replace
the key to be deleted. The \texttt{dellrg} helper function deletes the
largest key in a binary search tree, and produces a tuple of that key
and the new tree.

\clearpage

\begin{verbatim}
-- deletion without balancing
delete :: Ord a => a -> Bst a -> Bst a
delete _ E = E
delete k (T E k rt) = rt
delete k (T lt k E) = lt
delete k (T lt kt rt)
 | k<kt      = T (delete k lt) kt rt
 | k>kt      = T a kt (delete kt rt)
 | otherwise = T lt' k' rt
 where (lt',k') = dellrg lt

dellrg (T lt kt E) = (lt,kt)
dellrg (T lt kt rt) = (T lt' kt rt, k')
 where (lt',k') = dellrg lt
\end{verbatim}

Deletion from an AA tree is more complex than insertion, and we
put all the rebalancing logic into the \texttt{adjust} function.
This does not have to be applied when one child of
the key to be deleted is empty,
as the other child is unchanged and must be at level one.
But it is applied to trees constructed from the result of recursive
applications, including in the \texttt{dellrg} helper function.

\begin{verbatim}
-- deletion with balancing
delete :: Ord a => a -> AAt a -> AAt a
delete _ E = E
delete k (T lvt E k rt) = rt
delete k (T lvt lt k E) = lt
delete k (T lvt lt kt rt)
 | k<kt      = adjust (T lvt (delete k lt) kt rt)
 | k>kt      = adjust (T lvt lt kt (delete k rt))
 | otherwise = adjust (T lvt lt' k' rt)
 where (lt',k') = dellrg lt

dellrg (T lvt lt kt E) = (lt,kt)
dellrg (T lv lt kt rt) = (T lv lt' kt rt, k')
 where (lt',k') = dellrg lt
\end{verbatim}

The code for \texttt{adjust} is complicated, and requires careful case
analysis to understand.  It is still shorter and clearer than deletion
for any other balanced binary search tree implementation.

First, some helper functions to make the
code a bit shorter and clearer.
\texttt{lvl} produces the level of the root of a AA tree,
and \texttt{sngl} tests whether the root is single or double.

\begin{verbatim}
lvl E = 0
lvl (T lvt _ _ _) = lvt

sngl E = False
sngl (T _ _ _ E) = True
sngl (T lvx _ _ (T lvy _ _ _)) = lvx > lvy
\end{verbatim}

\clearpage

How can the invariant be broken by recursive deletion?
A child node may have its level lowered.
In general, a node being adjusted may have
one child two levels lower (as the result of the child being
constructed using the result of a recursive deletion), which violates
one of the invariants AA1 or AA2. 
We might need to drop the level of such a node to
restore that invariant.

We are now ready to describe \texttt{adjust}.
The easiest case is when each child of the argument tree \texttt{t}
has level no lower than one less than the level of \texttt{t}.
In this case, no adjustment is needed.

\begin{verbatim}
adjust t@(T lvt lt kt rt) | (lvl lt >= lvt-1) && (lvl rt >= lvt-1) = t
\end{verbatim}

If the right child of \texttt{t} has level
two below \texttt{t}, there are two subcases,
depending on whether the left child of \texttt{t}
(\texttt{x} in the diagram) is single or double.
The easier subcase is when \texttt{x} is single.
We can drop the level of \texttt{t} by one, fixing the
violation of AA2, but this
would result in a violation of AA3, since \texttt{t}
would then be at the same level as its left child.
But a skew operation restores the invariant AA3.

\includegraphics[scale=0.6]{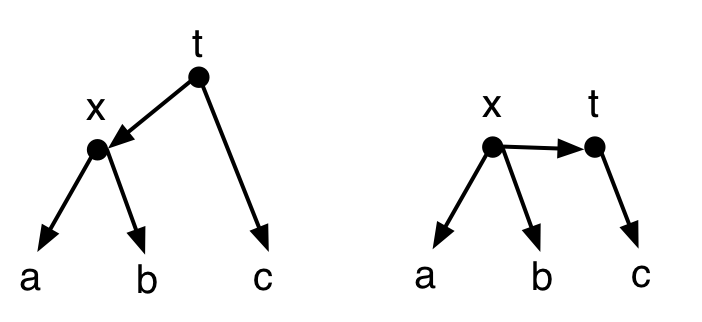}

The result is just a \texttt{skew} of \texttt{t}
with its level lowered by one.

\begin{verbatim}
adjust t@(T lvt lt kt rt) | (lvl rt < lvt-1) && sngl lt 
 = skew (T (lvt-1) lt kt rt)
\end{verbatim}

The harder subcase is when
the left child of \texttt{t} is double.
We need to do a more complicated restructuring.

\includegraphics[scale=0.6]{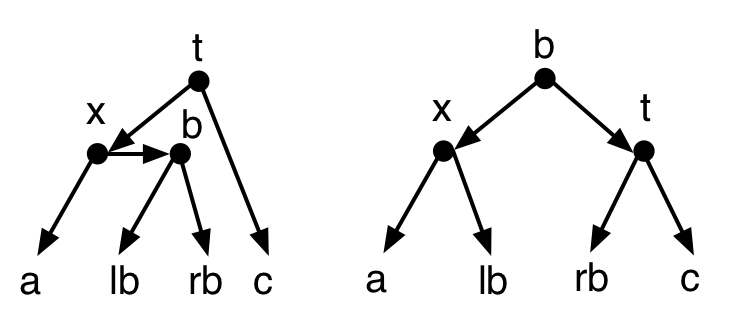}

\begin{verbatim}
adjust t@(T lvt lt kt rt) | (lvl rt < lvt-1)
   = let (T lvl a kl (T lvb lb kb rb)) = lt
     in T (lvb+1) (T lvl a kl lb) kb (T (lvt-1) rb kt rt)
\end{verbatim}

The remaining case of \texttt{adjust} is when the left child of \texttt{t}
has level two below \texttt{t}. Here again, there are subcases.

The first subcase is when \texttt{t} is single.
We can drop its level, but if its right child is double, this
will violate invariant AA3. But this is exactly what \texttt{split}
was designed to fix.

\begin{verbatim}
adjust t@(T lvt lt kt rt) | (lvl rt < lvt) = split (T (lvt-1) lt kt rt)
\end{verbatim}

\clearpage

The second subcase is when \texttt{t} is double,
so \texttt{t} and its right child \texttt{y}
are on the same level. 
We can drop the level of both \texttt{t} and \texttt{y} by one,
but then we have a mess,
because this might result in as many as five nodes on the same level
(\texttt{t}, \texttt{y}, \texttt{a}, \texttt{b},  and \texttt{d} in
the diagram below) and multiple violations of invariants. 
(In the diagram below, \texttt{a} is
drawn as double, but it might be single.)

\includegraphics[scale=0.6]{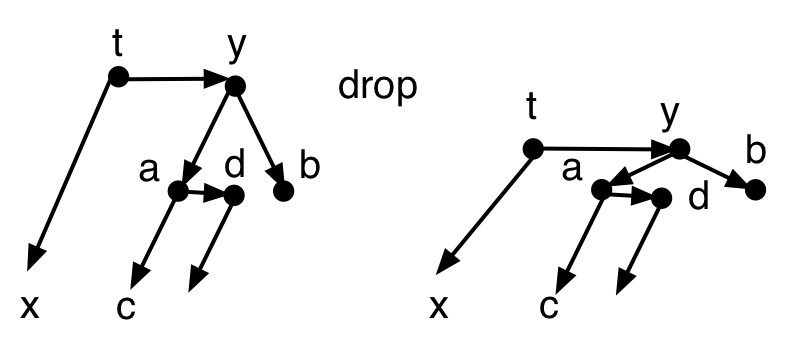}

We can explain the fixing up in terms of \texttt{skew}
and \texttt{split}, though we don't use them in the final code for
this case.
We start by naming the various values we need to manipulate,
based on the above diagram.

There are two sub-subcases of the second subcase, depending on whether \texttt{a} is
single or double. First, we consider the sub-subcase when \texttt{a} is single.
In this case, a \texttt{skew} followed by a \texttt{split} fixes
things up.

\includegraphics[scale=0.5]{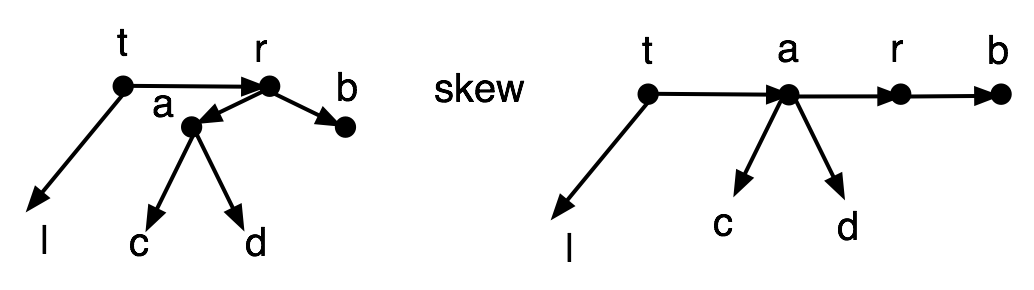}

\includegraphics[scale=0.5]{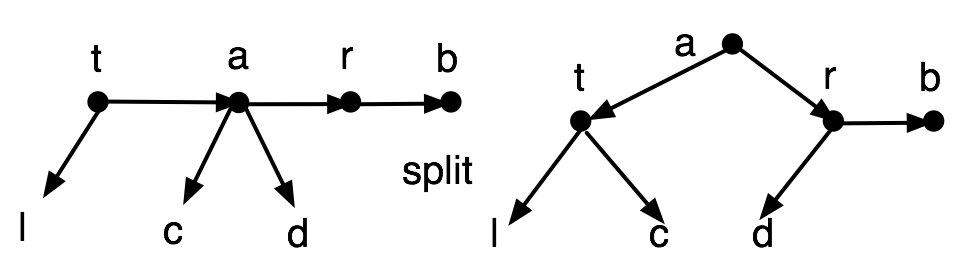}

If \texttt{b} was originally double, it might
need to be \texttt{split} at this point.

The sub-subcase where \texttt{a} is double is similar,
but two \texttt{skews} and two \texttt{splits} are needed, and
\texttt{r} ends up at a different level.

\includegraphics[scale=0.5]{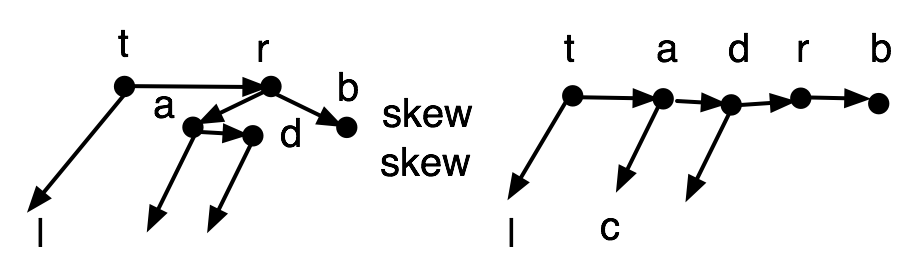}

\includegraphics[scale=0.6]{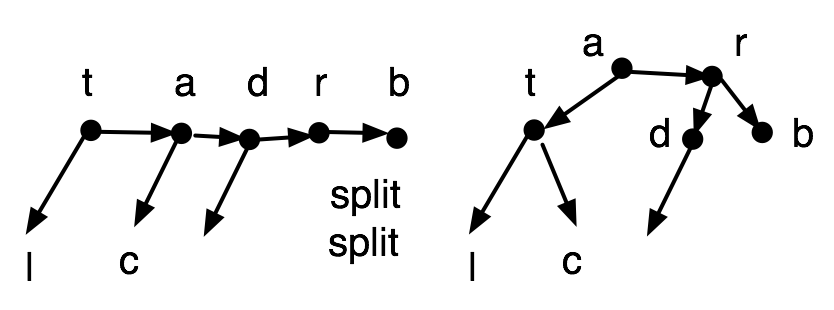}

The code for deletion in Andersson's paper applies mutating skew three
times and mutating split twice (sometimes to different nodes, so it
cannot be written as function composition). His implementation collapses some
cases to yield shorter code, at the cost of
understanding. Rather than do that here, let's take a look at the end
results of the single and double sub-subcases.

\includegraphics[scale=0.5]{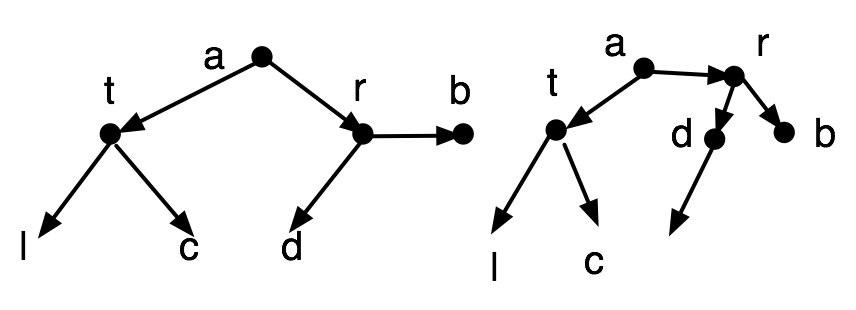}

These differ only in the level of \texttt{r}.
So we compute the new level accordingly,
using the \texttt{nlvl} helper function.

\begin{verbatim}
adjust t@(T lvt lt kt rt) | otherwise 
   = let (T lvr a@(T lva c ka d) kr b) = rt
     in T (lva+1) (T (lvt-1) lt kt c) ka (split (T (nlvl a) d kr b))

nlvl t@(T lvt _ _ _) = if sngl t then (lvt - 1) else lvt
\end{verbatim}

Here is the complete code for \texttt{adjust},
which is just the proper assemblage of the code
fragments presented above.

\begin{verbatim}
adjust t@(T lvt lt kt rt)
 | (lvl lt >= lvt-1) && (lvl rt >= lvt-1) = t
 | (lvl rt < lvt-1) && sngl lt = skew (T (lvt-1) lt kt rt)
 | (lvl rt < lvt-1) 
    = let (T lvl a kl (T lvb lb kb rb)) = lt
      in T (lvb+1) (T lvl a kl lb) kb (T (lvt-1) rb kt rt)
 | (lvl rt < lvt) = split (T (lvt-1) lt kt rt)
 | otherwise
   = let (T lvr a@(T lva c ka d) kr b) = rt
     in T (lva+1) (T (lvt-1) lt kt c) ka (split (T (nlvl a) d kr b))
\end{verbatim}

\section{Conclusions}

I use this presentation in a first ``advanced level'' course on
computer science for students with good mathematical skills (not
necessarily with prior computing experience). It is preceded by an
early treatment of Braun trees \cite{BR} (which implement sequences and are the
simplest useful data structure demonstrating logarithmic-depth trees)
and of naive elementary binary search trees. When we return to the
subject late in the term, we first discuss 2-3 and 2-3-4 trees, and
insertion into red-black trees based on Okasaki's version. Deletion
from AA trees is the last topic in the course.

Despite the fact that deletion from AA trees is simpler to explain and
implement than deletion from any other flavour of balanced binary
search trees, I would not recommend that it be covered in a first
course taken by all computer science majors, or by
non-majors. Since my course has non-majors enrolled, and I treat the
topic so late, there is no assignment based on the material, and
students are not responsible for it on the final exam. Consequently,
it serves mainly to close the gap and fulfil the promise of an
efficient implementation of maps, and as an advertisement for
the more complicated invariants of data structures to be seen in later
CS courses.

However, this topic fits well into a second-course treatment of
balanced binary search trees, whether for majors or non-majors.  Even
in a course using a conventional imperative garbage-collected language
for code examples, an initial immutable, recursive, functional
treatment will improve students' understanding of the topic prior to
covering the sort of code found in most textbooks and in library
implementations.

\section{Bibliography}

\nocite{*}
\bibliographystyle{eptcs}
\bibliography{tfpie}
\end{document}